\documentclass{aip-cp}

\usepackage[numbers]{natbib}
\usepackage{graphicx}

\usepackage{color}

\def\sumint{\setbox0=\hbox{$\displaystyle\sum$}\mathop{\rlap{\copy0}%
  \kern0pt \hbox to \wd 0{\hss$\displaystyle\int$\hss}}}

\begin{document}

\title{Electromagnetic Excitations and Responses in Nuclei from First Principles}

\author[aff1]{R. B. Baker\corref{cor1}}
\author[aff1]{K. D. Launey}
\author[aff2]{N. Nevo Dinur}
\author[aff3,aff2,aff4]{S. Bacca}
\author[aff1]{J. P. Draayer}
\author[aff1,aff5]{T.~Dytrych}

\affil[aff1]{Department of Physics and Astronomy, Louisiana State University, Baton Rouge, LA 70803, USA}
\affil[aff2]{TRIUMF, 4004 Wesbrook Mall, Vancouver, BC, V6T 2A3, Canada}
\affil[aff3]{Institut f\"ur Kernphysik and PRISMA Cluster of Excellence, Johannes Gutenberg-Universit\"at Mainz, 55128 Mainz, Germany}
\affil[aff4]{Department of Physics and Astronomy, University of Manitoba, Winnipeg, MB, R3T 2N2, Canada}
\affil[aff5]{Nuclear Physics Institute, Academy of Sciences of the Czech Republic, 250 68 \u{R}e\u{z}, Czech Republic}
\corresp[cor1]{Corresponding author: rbake25@lsu.edu}

\maketitle

\begin{abstract}
We discuss the role of clustering on monopole, dipole, and quadrupole excitations in nuclei in the framework of the \textit{ab initio} symmetry-adapted no-core shell model (SA-NCSM). The SA-NCSM starts from nucleon-nucleon potentials and, by exploring symmetries known to dominate the nuclear dynamics, can reach nuclei up through the calcium region by accommodating ultra-large model spaces critical to descriptions of clustering and collectivity. The results are based on calculations of electromagnetic sum rules and discretized responses using the Lanczos algorithm, that can be used to determine response functions, and for $^4$He are benchmarked against exact solutions of the hyperspherical harmonics method. In particular, we focus on He, Be, and O isotopes, including giant resonances and monopole sum rules.
\end{abstract}

\section{INTRODUCTION}
Nuclear response functions can provide valuable information about clustering, particularly by elucidating giant and pygmy resonances. For many nuclei, excitation strengths (e.g. monopole, dipole, etc.) can become fragmented, resulting in multiple peaks in the response function. These have been the focus of experiments for some time, e.g.\ see \cite{Youngblood1977, Youngblood1998, Lui, Gupta}, and they have provided ample opportunity for various theoretical approaches to successfully study the underlying cluster structure, e.g.\ see \cite{KE_Be, KE_C, Yamada, Kawabata, Chiba, Kimura1, Kimura2, Suzuki2}. Here, we seek to determine if these features also emerge from first principles. To achieve this, we utilize the \textit{ab initio} symmetry-adapted no-core shell model \cite{Launey, Dytrych}, which provides solutions in terms of a manageable number of collective and physically relevant basis states. Furthermore, in this framework, we are able to determine the individual contribution of each intrinsic deformation in these resonances.

\section{THEORETICAL FRAMEWORK}

\subsection{Symmetry-Adapted No-Core Shell Model}
The symmetry-adapted no-core shell model (SA-NCSM) utilizes emergent symmetries in nuclear physics to account for important collective correlations in nuclei, which allows one to reduce the computational complexity of nuclear structure calculations and, hence, reach model spaces that are currently unfeasible. It uses a collective basis describing deformation, plus rotations and vibrations thereof. Each basis state in this scheme is labeled schematically as $ |\vec{\gamma}\, N(\lambda\,\mu)\kappa L;(S_{p}S_{n})S;J M\rangle$. The SU(3) quantum numbers $(\lambda\ \mu)$ are associated with deformation, where, e.g.\ $(0\ 0)$, $(\lambda\ 0)$, and $(0\ \mu)$ describe spherical, prolate, and oblate shapes, respectively, $N$ is the total number of harmonic oscillator (HO) excitation quanta, and $L$ is the orbital angular momentum ($\kappa$ is multiplicity). The additional quantum numbers  $\vec{\gamma}$ are needed to distinguish among configurations carrying the same $N(\lambda\,\mu)$ and
($S_{p}S_{n})S$ labels. In this way, a complete shell-model basis is classified~\cite{Launey}. In the SA-NCSM, one can down-select the model space to only the physically relevant basis states, and typical SA-NCSM model spaces include the complete model space (all basis states) up to $N_{\rm max}=4$ or 6, and selected basis states extended to a much larger $N_{\rm max}$ (the SA-NCSM model spaces are hence denoted as $\left < 4 \right > N_{\rm max}$ or $\left < 6 \right > N_{\rm max}$). Here, we employ the SA-NCSM in an SU(3)-coupled basis to examine light nuclei up to oxygen isotopes. Using a realistic nucleon-nucleon $NN$ potential, we first calculate the many-body Hamiltonian and then we find the eigenenergies and many-body wave functions via the Lanczos algorithm. An important feature of the SA-NCSM is that the center-of-mass (CM) motion can be factored out exactly~\cite{Dytrych}. This ensures the translational invariance of the SA-NCSM wave functions.

\subsection{Lorentz Integral Transform Method}
To study the nuclear response of an external probe, we use the Lorentz integral transform (LIT) method \cite{Efros1, Efros2}. An electromagnetic response function is defined as 
\begin{equation}
R(\omega) = \sumint_f \left | \left < \psi_f \left | \hat{O} \right | \psi_0 \right > \right |^2 \delta(E_f - E_0 - \omega),
\label{eqn:response}
\end{equation}
where the sum runs over discrete and continuum final states of energy $E_f$, $\hat{O}$ is the excitation operator associated with the probe, $\omega$ is the energy transferred by the external probe, $\left | \psi_0 \right >$ and $\left | \psi_f \right >$ are the ground and excited states, respectively, and $E_0$ is the energy of the ground state~\cite{Efros2}. Given that the response function requires information about the ground state and all the excited states the excitation operator can connect to, direct calculations of $R(\omega)$ can be challenging. The Lorentz Integral Transform (LIT) method instead considers the quantity
\begin{equation}
L(\sigma, \Gamma) = \frac{\Gamma}{\pi} \int \mathrm{d}\omega \frac{R(\omega)}{(\omega - \sigma)^2 + \Gamma^2},
\label{eqn:LIT}
\end{equation}
which smears the response function with a Lorentzian peaked at $\sigma$ that has width $\Gamma$. Inserting the response function as defined above and invoking the closure relation, we can write $L(\sigma, \Gamma)$ in terms of the solution to a Schr\"odinger-like equation that is much simpler to solve. Typically, in the LIT method, one calculates the transform in Eq.~(\ref{eqn:LIT}) at large $\Gamma$ values until convergence is reached in the model space expansion and then one inverts it to obtain the response function. However, in the limit $\Gamma \rightarrow 0$, the Lorentzian becomes a delta function
\begin{equation}
L(\sigma, \Gamma \rightarrow 0) = \int \mathrm{d} \omega\ R(\omega) \delta (\omega - \sigma) = R(\sigma),
\end{equation}
and thus for small $\Gamma$, we obtain the discretized response function \cite{Efros2, Miorelli}. While the discretized response function is calculated with bound-state boundary conditions, it is useful to look at it if one wants to study where the discretized excited states are located in a given model space. When we calculate the discretized response function for small $\Gamma$ and we do not perform an inversion, we call this approach the ``Lanczos response method"~\cite{Efros2}.

In this study, we focus on several moments of the response function, also called sum rules, 
\begin{equation}
I_n = \int \mathrm{d} \omega \ \omega^n \ R(\omega),
\end{equation}
which, again with the help of the closure relation, one can rewrite as
\begin{equation}
I_n = \left < \psi_0 \left  | \hat{O}^\dagger\, \left( \mathcal{H}-E_0\right)^n\, \hat{O} \right | \psi_0 \right >.
\end{equation}
This means that such calculations depend on the given many-body Hamiltonian $\mathcal{H}$, along with its ground state energy and wave function, and the operator $\hat{O}$. Namely, it is a ground state expectation value of a new operator. Thus, the use of a bound-state basis is justified for its computation.

\subsection{Merging SA-NCSM and Lanczos response method}
Our procedure for calculating the response function with \textit{ab initio} SA-NCSM wave functions involves the following steps. First, we employ the {\it ab initio} SA-NCSM, with a given realistic interaction, to find the ground state wave function for the nucleus of interest. The SA-NCSM results, in particular the binding energy and rms radius of the ground state, are ensured to converge with respect to the maximum number of HO excitation quanta, $N_{\mathrm{max}}$. For a chosen excitation operator $\hat{O}$, we then construct a normalized pivot vector $\left | v_1 \right >$, given by
\begin{equation}
\left | v_1 \right >  = \frac{\hat{O} \left | \psi_0 \right >}{\sqrt{\left < \psi_0 \left | \hat{O}^{\dagger} \hat{O} \right | \psi_0 \right >}},
\label{eqn:pivot}
\end{equation}
where for $\hat{O}$ we use the isoscalar monopole, isovector dipole, and isoscalar quadrupole operators defined in the usual ways \cite{KE_Be, KE_C, Bahri}. The square of the quantity in the denominator is referred to as the non-energy weighted sum rule and is an important probe of the response function, as it represents the total transition strength from the ground state to all possible excited states.

From here, we use the same many-body Hamiltonian as the one used in the SA-NCSM calculations described above, and initiate the Lanczos algorithm with the pivot vector in Eq.~(\ref{eqn:pivot}) as our starting vector. The resulting Lanczos coefficients, i.e.\ the matrix elements of the tridiagonal matrix the Lanczos algorithm constructs, can then be used to calculate the transform $L(\sigma,\Gamma)$ and sum rules, as described in Ref.~\cite{Efros2, NevoDinur}.

This combined approach has two notable advantages: 1) By utilizing the SA-NCSM's ability to select only physically relevant basis states, we can perform larger calculations than the traditional NCSM and thus include important contributions from higher harmonic oscillator shells and reach heavier nuclei. 2) Within the SA-NCSM framework, we can decompose the wave function and examine the contribution of each individual basis state and its associated intrinsic deformation. This ability carries over to the response, where we can examine the contribution of each individual deformation to peaks in the response and identify emergent patterns. While other NCSM approaches have truncation techniques to accommodate large model spaces, this latter ability is novel. We also note that the SA-NCSM is not only a truncation scheme but a unique {\it ab initio} approach that reproduces nuclear collectivity up through the Ca region without effective charges~\cite{Launey_SOTANCP4}.

\section{RESULTS}

\subsection{Benchmark results for $^4$He}
\begin{figure}[h]
  \begin{tabular}{cc}
    \includegraphics[width=0.4\textwidth]{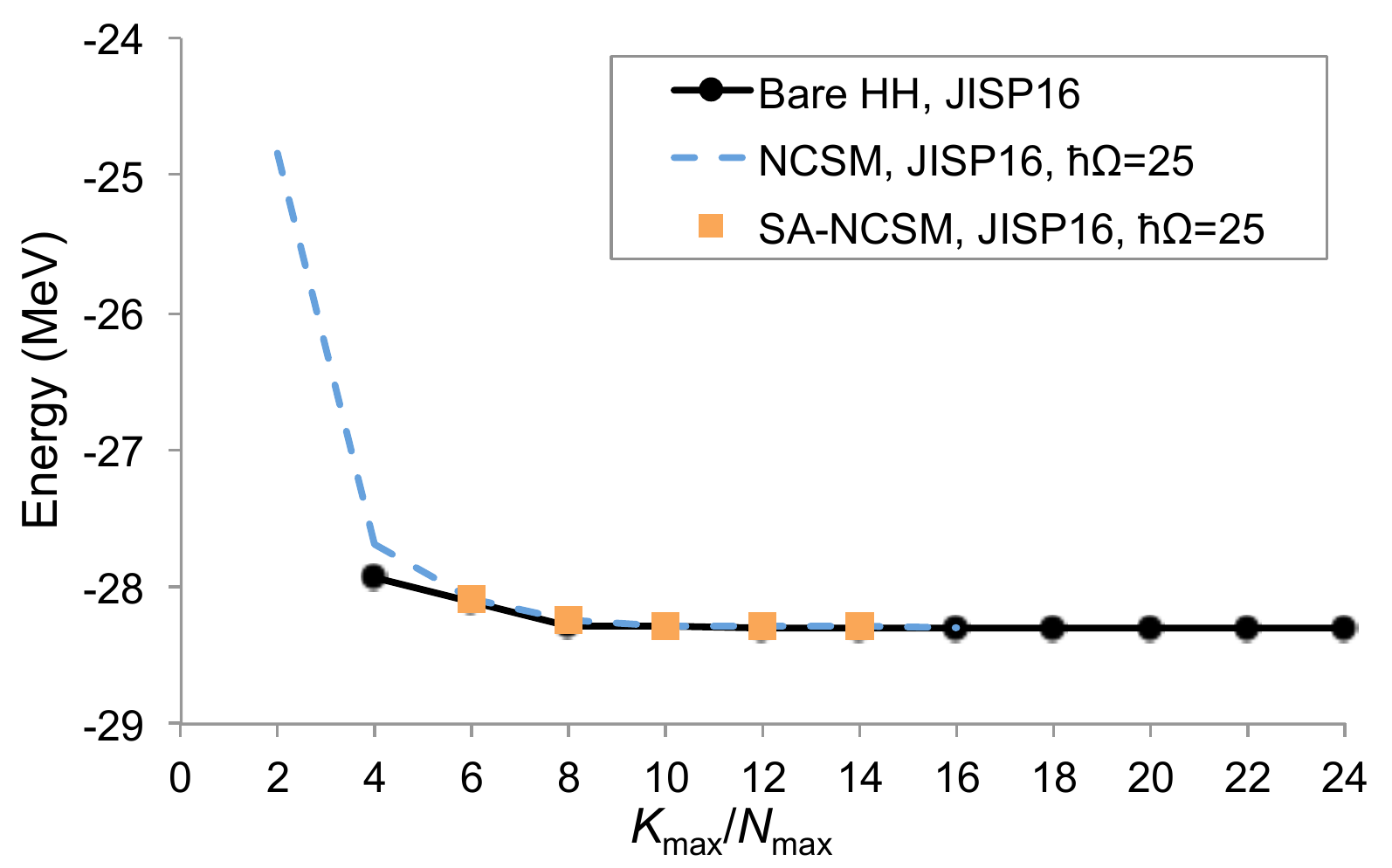} & \includegraphics[width=0.4\textwidth]{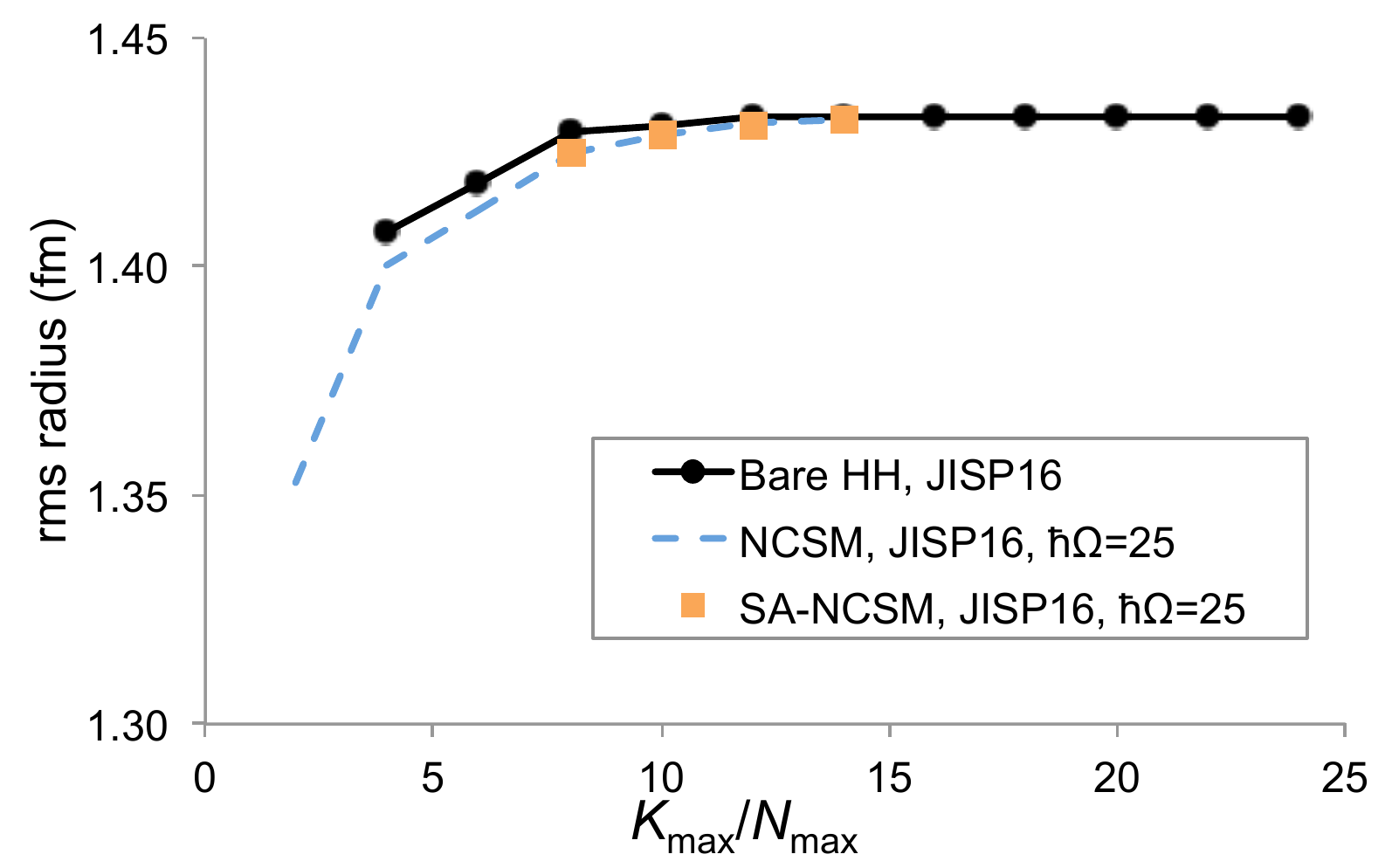}
  \end{tabular}
  \caption{(Left) Ground state energy and (right) rms radius of $^4$He as a function of $K_{\mathrm{max}}$ or $N_{\mathrm{max}}$ for bare HH, conventional NCSM, and SA-NCSM using the JISP16 interaction.}
  \label{fig:4He_gs}
\end{figure}

\begin{figure}[h]
  \centerline{\includegraphics[width=0.4\textwidth]{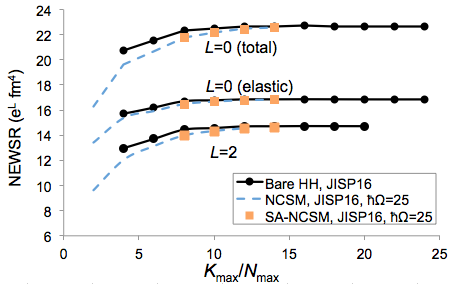}}
  \caption{Elastic and total monopole ($L=0$) and quadrupole ($L=2$) non-energy weighted sum rule (NEWSR) for $^4$He as a function of $K_{\mathrm{max}}$ or $N_{\mathrm{max}}$ for bare HH, conventional NCSM, and SA-NCSM using the JISP16 interaction.}
  \label{fig:4He_NEWSR}
\end{figure}

As a benchmark study, we calculate the ground state wave function for $^4$He from SA-NCSM compared to hyperspherical harmonics (HH), which is an exact method. The HH truncates its model space in terms of $K_{\mathrm{max}}$, the hyperspherical momentum, and is generally limited to smaller systems due to the use of Jacobi coordinates \cite{Efros2,Barnea}. Figure~\ref{fig:4He_gs} shows comparisons between the ground state properties of $^4$He, specifically its energy and rms radius, as calculated by HH, conventional NCSM or equivalently, the complete-space SA-NCSM, and SA-NCSM with SU(3) selection as described in Ref.\ \cite{Launey}, using the JISP16 realistic interaction. As can be seen there, all three methods are in good agreement with each other. Further, Fig.\ \ref{fig:4He_NEWSR} shows calculations for the non-energy weighted sum rule, i.e.\ ${\left < \psi_0 \left | \hat{O}^{\dagger} \hat{O} \right | \psi_0 \right >}$, for both the monopole and quadrupole operator. Again, all three methods are in good agreement.

\subsection{Response functions for $^{16}$O}
\begin{figure}[h]
  \begin{tabular}{cc}
    \includegraphics[width=0.4\textwidth]{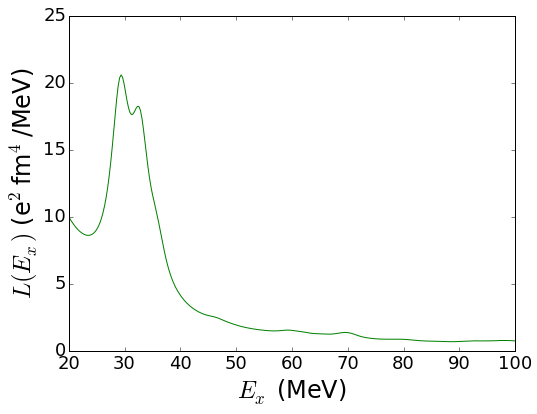} & \includegraphics[width=0.4\textwidth]{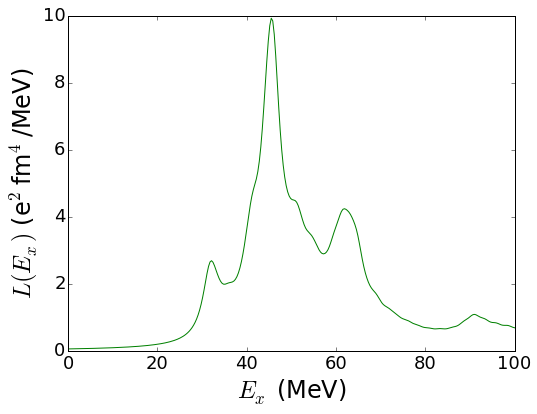}
  \end{tabular}
  \caption{(Left) Monopole and (right) quadrupole response for $^{16}$O for a $\left < 4 \right >8$ model space in the SA-NCSM using the JISP16 interaction, $\Gamma=2$ MeV, and $\hbar \Omega = 15$ MeV. Note that both plots qualitatively agree with the results from the softer interaction used in Ref.~\cite{Stumpf}.}
  \label{fig:16O}
\end{figure}

Using the SA-NCSM, we can apply the Lanczos response method to light and medium-mass nuclei. Figure \ref{fig:16O} shows the response functions with SA-NCSM wave functions for $^{16}$O with isoscalar monopole and quadrupole excitations using the JISP16 interaction and $\Gamma=2$ MeV. Utilizing the underlying framework of the SA-NCSM allows us to examine the intrinsic shapes contributing to each of these peaks in the response function. For the monopole response, two main peaks appear (aside from the elastic peak, which dominates at lower energies): more than $50\%$ of the lower energy peak is comprised of spin-0 configurations of $(\lambda\ \mu)=(2\ 0)$ and $(4\ 2)$, and the main contributions ($\sim 30\%$) to the higher energy peak are a spin-0 configuration of $(2\ 0)$ and spin-2 configuration of $(4\ 2)$.

For the quadrupole response, more than $25\%$ of the lowest energy peak is comprised of $(4\ 2)$, spin-0, whereas the main peak is composed $(\sim 30\%)$ of $(2\ 0)$ spin-0 and $(4\ 2)$ spin-2 configurations. Notably, since the ground state of $^{16}$O is dominated by a spherical configuration in the SA-NCSM, $(0\ 0)$, an excited $0^+ (2^+)$ state that is strongly connected to the ground state by a monopole (quadrupole) transition is expected to be dominated by $(2\ 0)$ deformation \cite{Launey}. Indeed, the operator known to generate 1p-1h giant monopole (quadrupole) excitations is exactly of $(2\ 0)$ SU(3) rank \cite{Bahri}. Similarly, the $(4\ 2)$, spin-0 configuration is a deformed 2p-2h configuration known to dominate the third $0^+$ in $^{16}$O \cite{Rowe}.

Moreover, as shown by Suzuki \cite{Suzuki}, the $(0\ 0)$, $(2\ 0)$, and $(4\ 2)$ components in $^{16}$O have a very high overlap with a cluster wave function of $^{12}$C and an alpha particle in their ground state with an associated relative motion. For further discussion, see Ref.~\cite{Launey_SOTANCP4}.

\subsection{Open-shell nucleus: $^{10}$Be}
\begin{figure}[h]
  \begin{tabular}{cc}
    \includegraphics[width=0.4\textwidth]{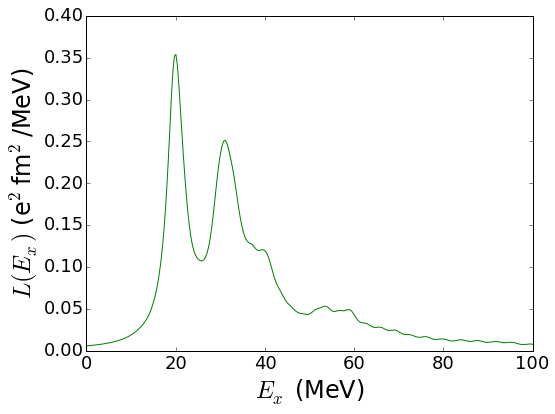} & \includegraphics[width=0.4\textwidth]{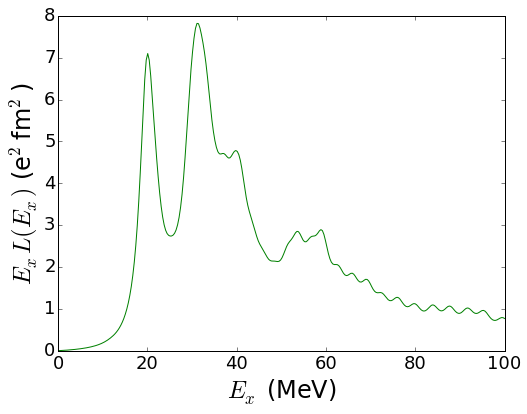}
  \end{tabular}
  \caption{(Left) Dipole response and (right) energy-weighted dipole response for $^{10}$Be for a $\left < 4 \right >10$ model space in the SA-NCSM using the JISP16 interaction, $\Gamma = 2$ MeV, and $\hbar\Omega=20$ MeV.}
  \label{fig:10Be}
\end{figure}

As an example of an open-shell nucleus within the reach of this approach, Fig.\ \ref{fig:10Be}a shows the isovector dipole response for $^{10}$Be. The ground state of $^{10}$Be is dominated by the configurations $(2\ 2)$ and $(3\ 0)$, and in the dipole response we find that the two largest peaks are largely composed of $(3\ 2)$, with the first peak being more than $35\%$ comprised of $(3\ 2)$, spin-0 and the second peak being more than $30\%$ a combination of $(3\ 2)$, spin-0 and spin-1.

Additionally, the results in Fig.~\ref{fig:10Be}b compare favorably, qualitatively, with the results for isovector dipole excitations in $^{10}$Be shown in Ref.~\cite{KE_Be}. Similar to these earlier results, the energy-weighted $E$1 strength also exhibits a two-peak structure and in both cases there is no contribution from the lowest $1^-$ state to this response function. A systematic study of the underlying physics as emerging within the SA-NCSM framework and detailed comparison to results of Ref.~\cite{KE_Be} can provide further insight into the role of clustering and collectivity, and is the focus of our near-future work.

\section{CONCLUSIONS}
We have produced a new first-principle approach to study the underlying shapes and dynamics inherent in nuclear responses. This technique combines the \textit{ab initio} SA-NCSM with the Lanczos response method, which together allows us to examine the peaks in the response function in terms of individual deformation contributions. This new approach has been benchmarked against the hyperspherical harmonics method for $^4$He and preliminary results for $^{16}$O and $^{10}$Be show qualitative agreement with existing literature. Additionally, analysis of the $^{16}$O results in this framework showed characteristic signs of giant monopole and quadrupole resonances. Future studies will work toward identifying these giant resonance characteristics from first principles in nuclei up through the Ca region.


\section{ACKNOWLEDGMENTS}
This work was supported in part by the U.S. NSF (OIA-1738287, ACI-1713690), SURA, and the Czech SF (16-16772S), and benefitted from high performance computational resources provided by LSU ({\tt www.hpc.lsu.edu}) and Blue Waters; the Blue Waters sustained-petascale computing project is supported by the National Science Foundation (awards OCI-0725070 and ACI-1238993) and the state of Illinois, and is a joint effort of the University of Illinois at Urbana-Champaign and its National Center for Supercomputing Applications.
A portion of the computational resources were provided by the National Energy Research Scientific Computing Center and by an INCITE award from the DOE Office of Advanced Scientific Computing.
Additional support was provided in part by the Natural Sciences and Engineering Research Council (NSERC), the National Research Council of Canada, by the Deutsche Forschungsgemeinschaft DFG through the Collaborative Research Center [The Low-Energy Frontier of the Standard Model (SFB 1044)], and through the Cluster of Excellence [Precision Physics, Fundamental Interactions and Structure of Matter (PRISMA)]. T. D. acknowledges support from Michal Pajr and CQK Holding.


\bibliographystyle{aipnum-cp}%
\bibliography{SOTANCP4.bib}%

\end{document}